\documentclass[twocolumn,amsmath,amssymb,prl,showpacs]{revtex4}
\usepackage{graphicx}
\begin{document}

\title{Controlling and enhancing THz collective electron dynamics in superlattices\\ 
by chaos-assisted miniband transport}
\author{M.T. Greenaway$^1$} 
\author{A.G. Balanov$^2$}
\author{E. Sch\"oll$^3$}
%\author{E. Scholl$^3$}
\author{T.M. Fromhold$^1$} 
\affiliation{$^1$School of Physics \& Astronomy, University of Nottingham, Nottingham NG7 2RD, United Kingdom \\
$^2$Department of Physics, Loughborough University, Leicestershire, LE11 3TU, United Kingdom \\
%$^3$Institut fur Theoretische Physik, Technische Universitat Berlin, 10623 Berlin, Germany}
$^3$Institut f\"ur Theoretische Physik, Technische Universit\"at Berlin, 10623 Berlin, Germany}

\begin{abstract}
We show that a tilted magnetic field transforms the structure and THz dynamics of charge domains in a biased semiconductor superlattice. At critical field values, strong coupling between the Bloch and cyclotron motion of a miniband electron triggers chaotic delocalization of the electron orbits, causing strong resonant enhancement of their drift velocity. This dramatically affects the collective electron behavior by inducing multiple propagating charge domains and GHz-THz current oscillations with frequencies ten times higher than with no tilted field.
\end{abstract}


\maketitle

Superlattices (SLs), comprising alternating layers of different semiconductor materials, provide a flexible environment for studying quantum transport in periodic potentials and for generating, detecting, mixing, and amplifying high-frequency electromagnetic radiation \cite{WAC2002,BON2005,SCHM1998, SUN1999, ELS2001,SCH98,RAS05,SHI03,SAV04,Endres,HYA2009}. Due to the formation of energy ``minibands'', electrons perform THz Bloch oscillations when a sufficiently high electric field, $F$, is applied along the SL. The Bloch orbits become more localized as $F$ increases, thus producing negative differential velocity (NDV) in the electron drift velocity, $v_d$, versus $F$ characteristic \cite{esaki-tsu,dohler}. This single-particle NDV also strongly influences the collective behavior of the electrons, causing them to slow and accumulate in high-density charge domains \cite{WAC2002,BON2005,SCHM1998}. For sufficiently high $F$, the domains are unstable and propagate through the SL, generating current oscillations at GHz-THz frequencies determined by the form of $v_d(F)$ and the SL length \cite{WAC2002,KAS1997,BON2002,AMA2003,HIZ2006}. %Control of the charge domains, and their oscillation frequency, has so far been achieved by using the SL layer structure to tailor the $v_d(F)$ curve, for example by creating several peaks and NDV regions \cite{?}, which generate current oscillations at distinct frequencies. 
In a given SL with fixed $v_d(F)$, the frequency of domain dynamics can be tuned, over a limited range, by changing the applied voltage \cite{KAS1997,SCH98}. 

\begin{figure}[!]%f1
  \centering
\includegraphics*[width=1.\linewidth]{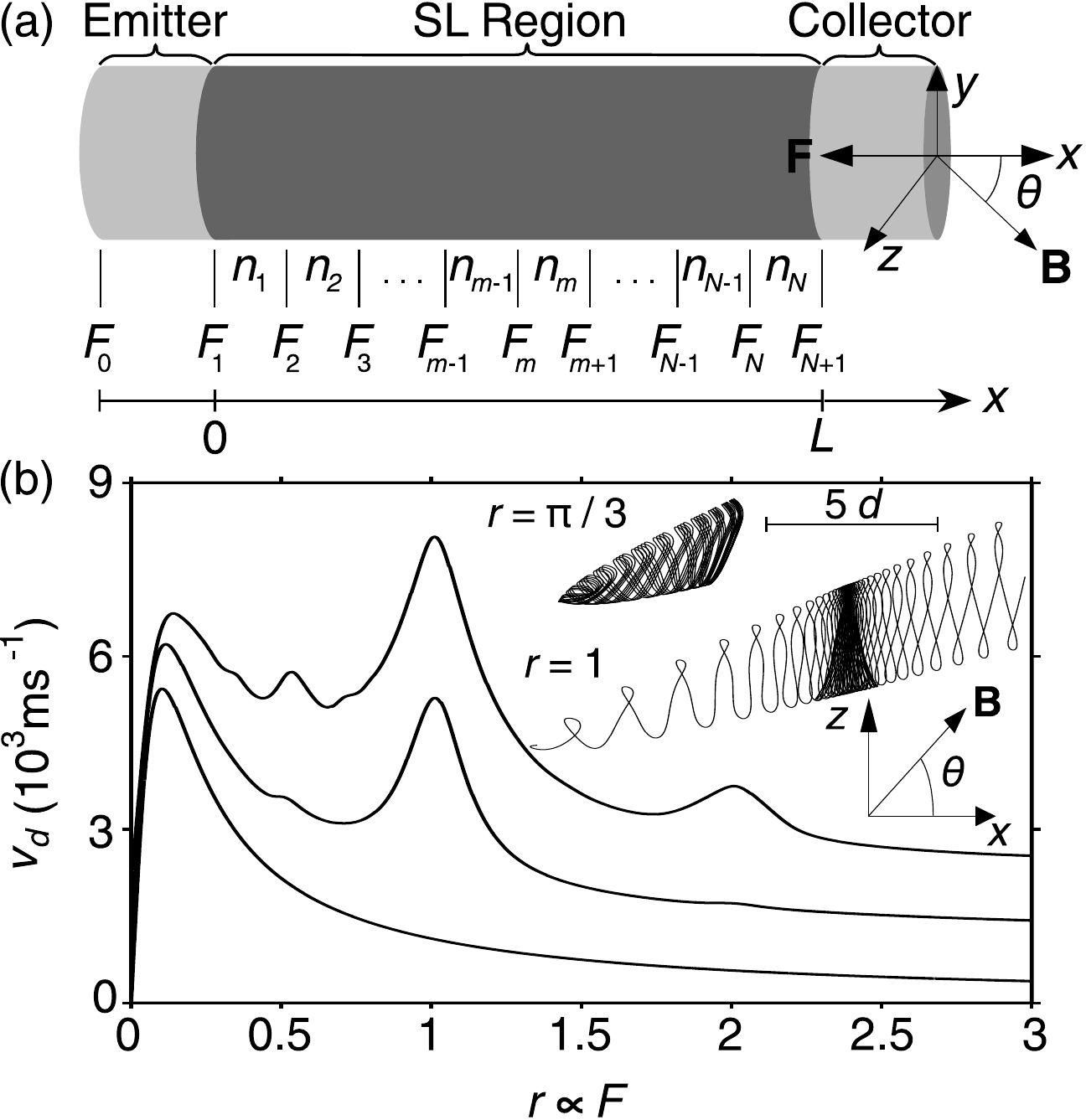}
\caption{(a) Semiclassical model in which the SL is described by a continuum region (dark gray) with miniband dispersion $E(k_x)$. We discretize this region into $N$ layers. In the $m^{th}$ layer, the electron density is $n_m$. At the left- [right-] hand edges of this layer, $F$ = $F_m$ [$F_{m+1}$]. Coordinate axes show orientation of \textbf{F} and \textbf{B}. (b) $v_d$ versus $r\propto F$ calculated for $B = 15$ T with (from bottom to top) $\theta = 0$, $ 25^{\circ}$ and $40^{\circ}$. For clarity, curves are offset vertically by $10^3$ ms$^{-1}$. Inset: electron trajectories in $x-z$ plane (scale bar at top) calculated over 4 ps for $r=1$ (lower) and $r=\pi /3$ (upper) at $\theta=40^{\circ}$. 
\label{fig:SL}}
\end{figure}

Here, we show that both the spatial profile of charge domains and their oscillation frequency can be flexibly controlled by using a tilted magnetic field, \textbf{B}, to engineer the $v_d(F)$ characteristic of the SL. At $F$ values for which the Bloch frequency equals the cyclotron frequency corresponding to the \textbf{B} component along the SL axis, the semiclassical electron motion changes abruptly from localized stable trajectories to unbounded chaotic paths, which propagate rapidly through the SL \cite{FRO2001,FRO2004,FOW2007,BAL2008,STA04}. This delocalization creates a series of sharp resonant peaks in $v_d(F)$, which were detected in previous DC current-voltage measurements \cite{FRO2004,FOW2007,BAL2008}, relate to mode coupling in Josephson junctions \cite{Kosevich}, and can stabilize the SL Bloch gain profile in the vicinity of Stark-cyclotron resonances \cite{Hyart09}. We show that these $v_d$ peaks create multiple propagating charge domains, shaped by $B$ and $\theta$, and thereby generate AC currents whose magnitude and frequencies are far higher than when $B=0$. Chaos-assisted single-electron transport induced by the interplay between cyclotron and Bloch motion therefore provides a mechanism for controlling the collective dynamics of the miniband electrons, thus increasing the power and frequency of the resulting current oscillations by an order of magnitude.

We consider the GaAs/AlAs/InAs SL used in recent experiments \cite{FRO2004,FOW2007}. Fourteen unit cells, each of width $d=8.3$ nm, form the SL, which is enclosed by GaAs ohmic contacts [light gray in Fig. \ref{fig:SL}(a)] with n-doping density $n_0=10^{23}$ m$^{-3}$. Electrons are confined to the first miniband with kinetic energy versus wavenumber, $k_x$, dispersion relation $E(k_x) =\Delta[1-\cos (k_x d)]/2$, where the miniband width $\Delta=19.1$ meV \cite{FRO2004}. Semiclassical miniband transport corresponds to modeling the SL by a region of width $L$ [dark gray in Fig. \ref{fig:SL}(a)] where electrons move freely (with the GaAs effective mass $m^{*}$) in the $y-z$ plane but have dispersion, $E(k_x)$, along the SL axis. 

We calculated semiclassical trajectories for a miniband electron with \textbf{B} tilted at an angle $\theta$ to the SL ($x$) axis [Fig. \ref{fig:SL}(a)] and $F$ uniform throughout the SL \cite{FRO2001,FRO2004}. We then used an Esaki-Tsu (ET) model \cite{esaki-tsu, FRO2004} to determine $v_d=\langle v_x (t)\exp(-t/\tau)\rangle/\tau_i$, where $v_x (t)$ is the electron speed along $x$ at time $t$ and $\langle .\rangle$ denotes averaging over the starting velocities of the hot miniband electrons, whose temperature ($\approx 100$ K) exceeds that of the lattice (4.2 K), and integration over $t$, taking an electron scattering time $\tau=\tau_i[\tau_e/(\tau_e + \tau_i)]^{1/2}=250$ fs determined from the elastic (interface roughness) scattering time $\tau_e$ = 29 fs and the inelastic (phonon) scattering time $\tau_i$ = 2.1 ps \cite{FOW2007,footA}. When $\theta=0^{\circ}$, cyclotron motion in the $y-z$ plane is separable from the Bloch motion along $x$. The lower curve in Fig. \ref{fig:SL}(b) shows $v_d$ versus $r = \omega_B / \omega_c \propto F$ where $\omega_B=eFd/\hbar$ is the Bloch frequency and $\omega_c=eB \cos \theta /m^{*}$ is the cyclotron frequency corresponding to the $x-$component of \textbf{B}. As expected \cite{esaki-tsu,dohler,WAC2002}, this trace peaks when $r=1/\omega_c\tau$ (i.e. $\omega_B \tau =1$) and thereafter decreases with increasing $r$ as more electrons complete Bloch orbits before scattering.

When $\theta \neq 0^{\circ}$, strong mixing of the cyclotron and Bloch motion drives the electron orbits chaotic [Fig. \ref{fig:SL}(b) inset] \cite{FRO2001,FRO2004,BAL2008}. %In this regime, the orbital width along $x$ depends critically on $r$. 
When $r$ is irrational, the electron orbits remain localized along $x$ [Fig. \ref{fig:SL}(b) upper inset]. By contrast, when $r$ is an integer the electrons follow unbounded paths [Fig. \ref{fig:SL}(b) lower inset] and map out intricate ``stochastic web'' patterns in phase space \cite{FRO2001,footB}. This abrupt delocalization of the electron paths generates sharp resonant peaks in $v_d$. The $v_d(r)$ curves shown in Fig. \ref{fig:SL}(b) for $\theta=25^{\circ}$ (middle trace) and $40^{\circ}$ (top trace) reveal a large peak at $r=1$ and a smaller additional feature at $r=2$, most apparent when $\theta=40^{\circ}$. For $r$ values that are rational but not integer, the electron orbits are finite, but exhibit some resonant extension along $x$. This causes the small additional peaks visible at $r=0.5$ in the middle and top curves of Fig. \ref{fig:SL}(b). It is well known that NDV in SLs and Gunn diodes creates propagating charge domains \cite{WAC2002}. The multiple NDV regions associated with chaos-assisted resonant transport when $\theta \neq 0^{\circ}$, suggests that the tilted \textbf{B}-field will induce more complex spatio-temporal domain dynamics.  

To investigate the collective behavior of the electrons, we solved the current-continuity and Poisson equations self-consistently throughout the device by adapting, for miniband transport, a model used previously to describe inter-well transitions in SLs \cite{WAC2002,BON2002,HIZ2006}. In this model, we discretize the miniband transport region [dark gray in Fig. \ref{fig:SL}(a)] into $N=480$ layers, each of width $\Delta x=L/N=0.24$ nm small enough to approximate a continuum. The volume electron density in the $m^{th}$ layer (with right-hand edge at $x=m \Delta x$) is $n_m$ and the $F$ values at the left- and right-hand edges of this layer [vertical lines in Fig. \ref{fig:SL}(a)] are $F_m$ and $F_{m+1}$ respectively. In the emitter and collector ohmic contacts, $F= F_0$. The evolution of the charge density in each layer is given by the current continuity equation
\begin{equation}
 e \Delta x \frac{d n_m}{d t} = J_{m - 1} - J_m, \ \ \ m = 1 \ldots N,	\label{eq:continuity} 
\end{equation}
\noindent
where $e>0$ is the electron charge and $J_m = e n_m v_d(\overline{F_m})$, in which $\overline{F_m}$ is the mean field in the $m^{th}$ layer \cite{foot1}, is the areal current density from the $m^{th}$ to the $m+1^{th}$ layer neglecting diffusion \cite{HIZ2006,FRO2004}. Since $J_m$ depends on the local drift velocity, $v_d(\overline{F_m})$, the collective electron dynamics depend directly on the single electron orbits. In each layer, $F_m$ obeys the discretized Poisson equation 
\begin{equation}
F_{m + 1} = \frac{e \Delta x}{\varepsilon_0 \varepsilon_r} \left( n_m - n_D \right) + F_m, \ \ \ m = 1 \ldots N,  \label{eq:poisson} \\
% e \frac{d n_m}{d t} = J_{m - 1} - J_m \ \ \ \textrm{for} \ \ \ m = 1 \ldots N	\label{eq:continuity} 
\end{equation}
\noindent
where $\varepsilon_0$ and $\varepsilon_r=12.5$ are, respectively, the absolute and relative permittivities and $n_D=3\times 10^{22}$ m$^{-3}$ is the n-type doping density in the SL layers \cite{FRO2004}.

We use ohmic boundary conditions \cite{WAC2002} to determine the current, $J_0 = \sigma F_0$, in the heavily-doped emitter of electrical conductivity $\sigma= $ 3788 Sm$^{-1}$ \cite{FRO2004}. The voltage, $V$, applied to the device is a global constraint given by $V = U + \frac{\Delta x}{2} \sum_{m = 1}^N (F_m + F_{m + 1})$, where the voltage, $U$, dropped across the contacts includes the effect of charge accumulation and depletion in the emitter and collector regions and a $17$ $\Omega$ contact resistance \cite{Trav_Phd}. We calculate the current $I(t) = \frac{A}{N+1} \sum_{m = 0}^N J_m^{}$,  where $A=5 \times 10^{-10}$ m$^2$ is the cross-sectional area of the SL \cite{FRO2004}.

\begin{figure}%f1
  \centering
\includegraphics*[width=.8\linewidth]{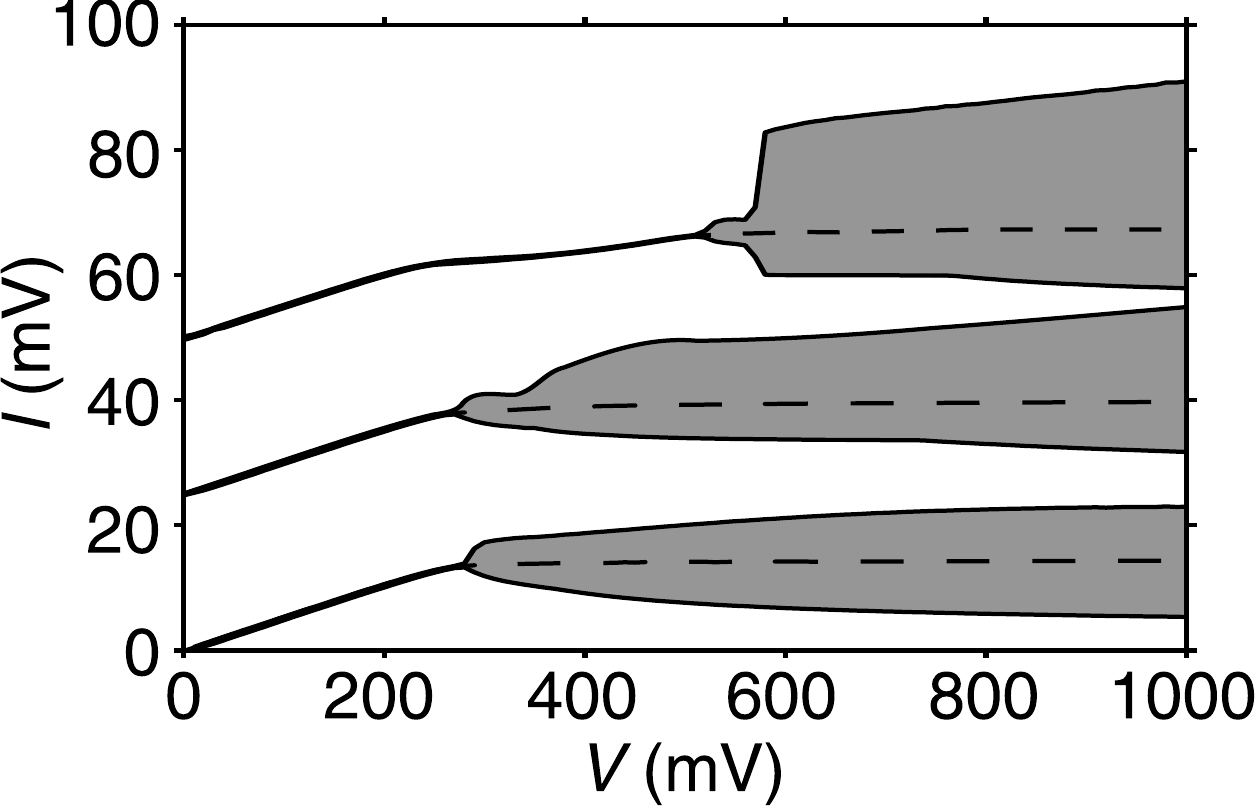}
  \caption{$I(V)$ characteristics calculated for (from bottom to top) $\theta=0^{\circ}$, $25^{\circ}$, and $40^{\circ}$. %For clarity, 
  Curves are vertically offset by $25$ mA. Current oscillations occur within the shaded regions, whose upper [lower] bounds are $I_{\max}(V)$ [$I_{\min}(V)$]. Dashed curves are unstable steady state solutions of $I$ \label{fig:IV}}
\end{figure}

Following initial transient behavior, $I(t)$ either reaches a constant value or oscillates between minima and maxima, $I_{\min}$ and $I_{\max}$ respectively, which depend on $V$, $B$ and $\theta$. Figure \ref{fig:IV} shows $I(V)$ curves calculated for $B=15$ T and (from bottom to top) $\theta=0^{\circ}$, $25^{\circ}$, and $40^{\circ}$. Each trace is single valued at low $V$, but double valued in the shaded regions, where the lower [upper] boundaries show $I_{\min}$ [$I_{\max}$]. For all $\theta$, stationary behavior occurs at low $V$ where $I(V)$ is approximately linear. But when $V$ exceeds a critical value, $V_c$, which depends on $B$ and $\theta$, the stationary state loses its stability via Hopf bifurcation and $I(t)$ starts to oscillate between $I_{\min}$ and $I_{\max}$. %In this regime, 
The amplitude of the oscillations, $I_a=I_{\max}-I_{\min}$, increases with increasing $V$ for all $\theta$. In addition, for given $V$, $I_a$ generally increases with increasing $\theta$. In the regime where $I(t)$ oscillates (shaded in Fig. \ref{fig:IV}), there is also an \emph{unstable} stationary state, found by setting $dn_m / dt=0$ in Eq. (1), corresponding to a fixed current whose $V$-dependence is shown by the dashed curves in Fig. \ref{fig:IV}. The shapes of the stationary $I(V)$ curves, each comprising a stable ($V<V_c$) and unstable ($V \geqslant V_c$) part, are similar to previous experimental measurements \cite{FRO2004}. 

\begin{figure}%f1
  \centering
\includegraphics*[width=1.\linewidth]{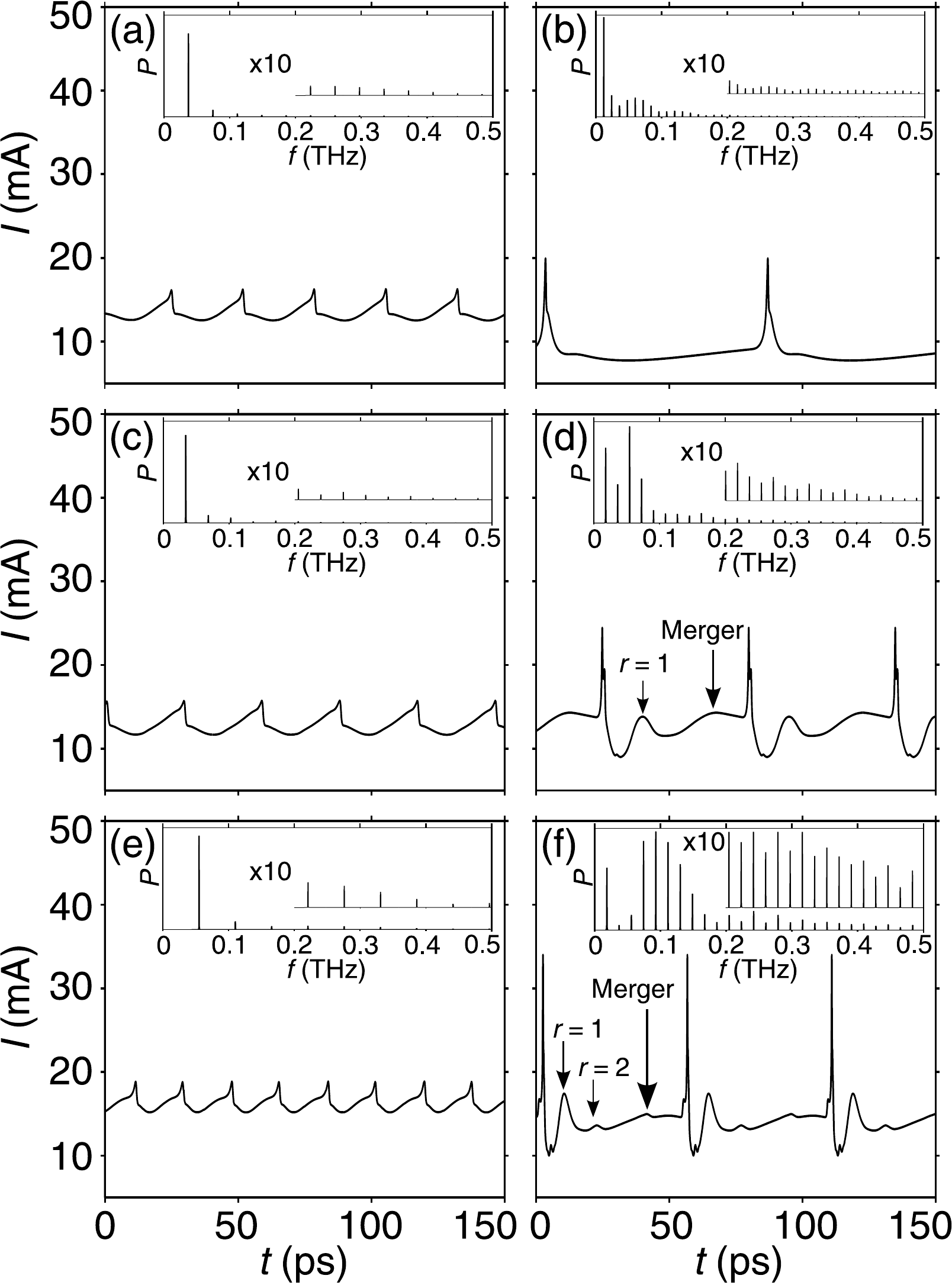}
  \caption{$I(t)$ curves calculated for [$\theta ,V$] = (a) [$0^{\circ}$, 290 mV]; (b) [$0^{\circ}$, 490 mV]; (c) [$25^{\circ}$, 290 mV]; (d) [$25^{\circ}$, 490 mV]; (e) [$40^{\circ}$, 540 mV]; (f) [$40^{\circ}$, 610 mV]. Arrowed peaks in (d) and (f) are discussed in text. Insets: Fourier power spectra, $P(f)$, with a common vertical scale in arb. units and, for $f \gtrsim$ 0.2 THz, also shown vertically enlarged $\times 10$ and offset. \label{fig:It}}
\end{figure}

\begin{figure}%f1
  \centering
\includegraphics*[width=1.\linewidth]{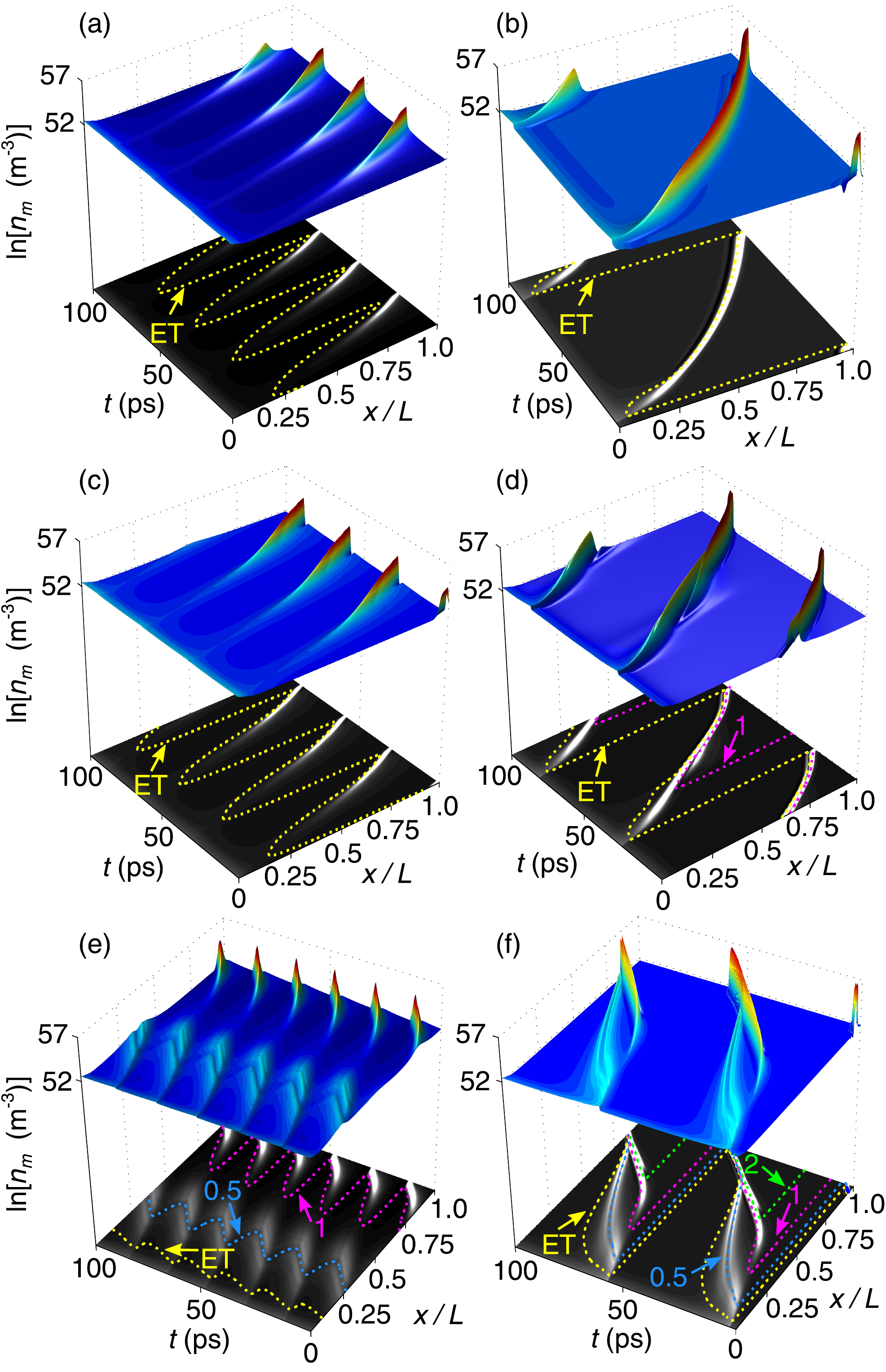}
  \caption{(Color) $n_m$ calculated for [$\theta, V$] = (a) [$0^{\circ}$, 290 mV]; (b) [$0^{\circ}$, 490 mV]; (c) [$25^{\circ}$, 290 mV]; (d) [$25^{\circ}$, 490 mV]; (e) [$40^{\circ}$, 540 mV]; (f) [$40^{\circ}$, 610 mV]. For clarity, upper (surface) plots are shown as grey-scale projections beneath where yellow, blue, purple and green curves are loci of constant $F$ values corresponding, respectively, to the ET, $r=0.5$, $r=1$ and $r=2$ $v_d(F)$ peaks. \label{fig:chargedensity}}
\end{figure}

We now consider how the $I(t)$ curves vary with $V$ and $\theta$. For $\theta=0^{\circ}$ and $V=290$ mV $\approx V_c$ [Fig. \ref{fig:It}(a)], $I(t)$ exhibits periodic oscillations whose frequency $\sim$ 37 GHz corresponds to the single dominant peak in the Fourier power spectrum, $P(f)$, inset. When $V$ increases to 490 mV [Fig. \ref{fig:It}(b)], the fundamental frequency of the oscillations falls to $\sim$ 12 GHz. In addition, the peaks in $I(t)$ sharpen, thus strengthening the higher frequency harmonics in $P(f)$ [Fig. \ref{fig:It}(b) inset]. The $I(t)$ and $P(f)$ curves calculated for $\theta=25^{\circ}$ and $V=290$ mV $\approx V_c$ [Fig. \ref{fig:It}(c)] are similar to those for $\theta=0^{\circ}$ [Fig. \ref{fig:It}(a)]. But when $V$ increases to 490 mV [Fig. \ref{fig:It}(d)], the results for $\theta=25^{\circ}$ differ markedly from those for $\theta=0^{\circ}$ [Fig. \ref{fig:It}(b)]. In particular, tilting \textbf{B} almost doubles the fundamental frequency and introduces new $I(t)$ peaks [arrowed in Fig. \ref{fig:It}(d)] whose origin we explain below. Compared with $\theta=0^{\circ}$, these extra peaks strongly enhance the high frequency components in $P(f)$, as shown in Fig. \ref{fig:It}(d) inset, which reveals a dominant $3^{rd}$ harmonic at 54 GHz. 

When $\theta=40^{\circ}$ and $V=540$ mV $\approx V_c$ [Fig. \ref{fig:It}(e)], the shapes of the $I(t)$ and $P(f)$ curves are similar to those for $\theta=0^{\circ}$ and $25^{\circ}$, but the dominant fundamental frequency is higher. By contrast, when $\theta=40^{\circ}$ and $V=610$ mV [Fig. \ref{fig:It}(f)], the $I(t)$ fluctuations are both stronger and richer than for $\theta=0^{\circ}$ [Fig. \ref{fig:It}(b)] and $\theta=25^{\circ}$ [Fig. \ref{fig:It}(d)]. Consequently, the high frequency peaks in $P(f)$ [Fig. \ref{fig:It}(f) inset] are greatly enhanced, with the $5^{th}$ harmonic at 92 GHz being the strongest.  

To understand how $I(t)$ varies with $V$ and $\theta$, we consider how these parameters affect the underlying spatio-temporal electron dynamics. Figure \ref{fig:chargedensity}(a) shows the surface plot of $n_m(t,x)$ with its grey-scale projection beneath calculated for $\theta=0^{\circ}$ and $V$ = 290 mV $\approx V_c$. For given $x$, $n_m$ oscillates periodically in $t$ due to NDV in the corresponding $v_d(r)$ curve [lower trace in Fig. 1(b)]. The dashed yellow curve in the lower projection of Fig. \ref{fig:chargedensity}(a) shows the $(t,x)$ locus along which $F$ is fixed at the value corresponding to the ET peak in Fig. \ref{fig:SL}(b) lower inset. As $x$ passes beyond this locus, the electrons slow, thus increasing the local values of both $n_m$ and $F$. This further decreases $v_d$ and increases $n_m$, making the electrons accumulate in a charge domain [peak in surface plot %light gray 
in Fig. \ref{fig:chargedensity}(a)], which propagates through the SL \cite{WAC2002}. When the domain reaches the collector ($x=L$) it produces a sharp increase in $I(t)$. Another charge domain then forms near the emitter and the propagation process repeats, so producing $I(t)$ oscillations \cite{WAC2002}. For larger $V$ [Fig. \ref{fig:chargedensity}(b)] there are similar domain dynamics, but their frequency is lower because the higher mean $F$ value reduces $v_d$.

When $V$ = 290 mV $\approx V_c$, increasing $\theta$ from $0^{\circ}$ [Fig. \ref{fig:chargedensity}(a)] to $25^{\circ}$ [Fig. \ref{fig:chargedensity}(c)] has little qualitative effect on the domain dynamics because $V$ is low enough to ensure $r \lesssim 0.7$ through most of the SL: a regime where the $v_d(r)$ curves for $\theta = 0^{\circ}$ and $25^{\circ}$ have similar shapes [Fig. \ref{fig:SL}(b)].

This picture changes qualitatively when $V$ becomes high enough to make $r\geqslant 1$ at some positions within the SL. Figure \ref{fig:chargedensity}(d) illustrates this for $V$ = 490 mV and $\theta = 25^{\circ}$. The yellow and purple curves in the lower projection show the $(t,x)$ loci along which $F$ equals the values corresponding, respectively, to the leftmost (ET) and $r=1$ $v_d$ peaks in Fig. \ref{fig:SL}(b). When $t \approx$ 25 ps, NDV associated with the ET peak creates a high density charge domain for $x$ just beyond the yellow locus. At $t \approx$ 50 ps, a second charge accumulation region appears above the purple locus. This domain originates from the NDV region just beyond the $r=1$ $v_d$ peak. Its appearance produces an additional peak, labeled ``$r=1$'', in the $I(t)$ trace in Fig. \ref{fig:It}(d). When $t \approx$ 65 ps, merger of the two charge domains creates the $I(t)$ peak labeled ``Merger'' in Fig. \ref{fig:It}(d). After merger, the charge within the single domain is almost twice that for $\theta = 0^{\circ}$. In addition, the presence of the $r=1$ $v_d$ peak increases the mean electron drift velocity compared with $\theta=0^{\circ}$, thus also raising the domain propagation speed. These two factors increase both the frequency and amplitude of the $I(t)$ oscillations [compare Figs. \ref{fig:It}(b) and (d) and their insets]. %The appearance of the ``Merger'' peak in $I(t)$ further strengthens the high-frequency harmonics in the $P(f)$ [compare insets to Figs. \ref{fig:It}(b) and (d)].  

%However, increasing $V$ to ?? mV produces a qualitative change in the domain dynamics [Fig.\ref{fig:chargedensity}(d)]. In particular, the rate  at which the domains reach the collector is ~ ?? times higher than at the $V$ value when $\theta=0^{\circ}$, meaning that the frequency of the corresponding $I(t)$ oscillations is also higher [compare Figs. \ref{fig:It}(b) and (d)]. Increasing $\theta$ to $40^{\circ}$ further increases the domain oscillation frequency both near the Hopf bifurcation [Fig.\ref{fig:chargedensity}(e)] and at higher $V=$ ?? mV [Fig.\ref{fig:chargedensity}(f)]. The tilted \textbf{B} field also induces additional high-density charge domain filaments, which are arrowed in Figs.\ref{fig:chargedensity}(d)-(f). These filaments, which are particularly striking when $\theta$ increases to $40^{\circ}$, cause the increased domain speed as we now explain. \textbf{Need to relate domain diagrams to Trav's static plots}.

Increasing $\theta$ to $40^{\circ}$ further enriches the charge domain patterns. Since the $r=0.5,1$ and 2 resonances occur for smaller $F$ at higher $\theta$, their effect on the domain dynamics is apparent even for $V$ very close to $V_c$. Figure \ref{fig:chargedensity}(e) reveals multiple charge domains near the yellow, blue and purple loci in the lower projection, along which $F$ coincides, respectively, with the ET, $r=0.5$ and $r=1$ $v_d$ peaks [upper curve in Fig. \ref{fig:SL}(b)]. Coexistence of multiple domains substantially increases both the amplitude and frequency of the $I(t)$ oscillations [compare Figs. \ref{fig:It}(a) and (e)]. When $V \approx$ 610 mV [Fig. \ref{fig:chargedensity}(f)] a new domain associated with the $r=2$ resonance (green locus) appears. The various domains produce multiple peaks in $I(t)$, as shown in Fig. \ref{fig:It}(f) where the labels mark peaks arising from formation of the $r=1$ and $2$ domains and their merger. These peaks create strong high-frequency components in $P(f)$ [Fig. \ref{fig:It}(f) inset]. 

\begin{figure}%f1
  \centering
\includegraphics*[width=1.\linewidth]{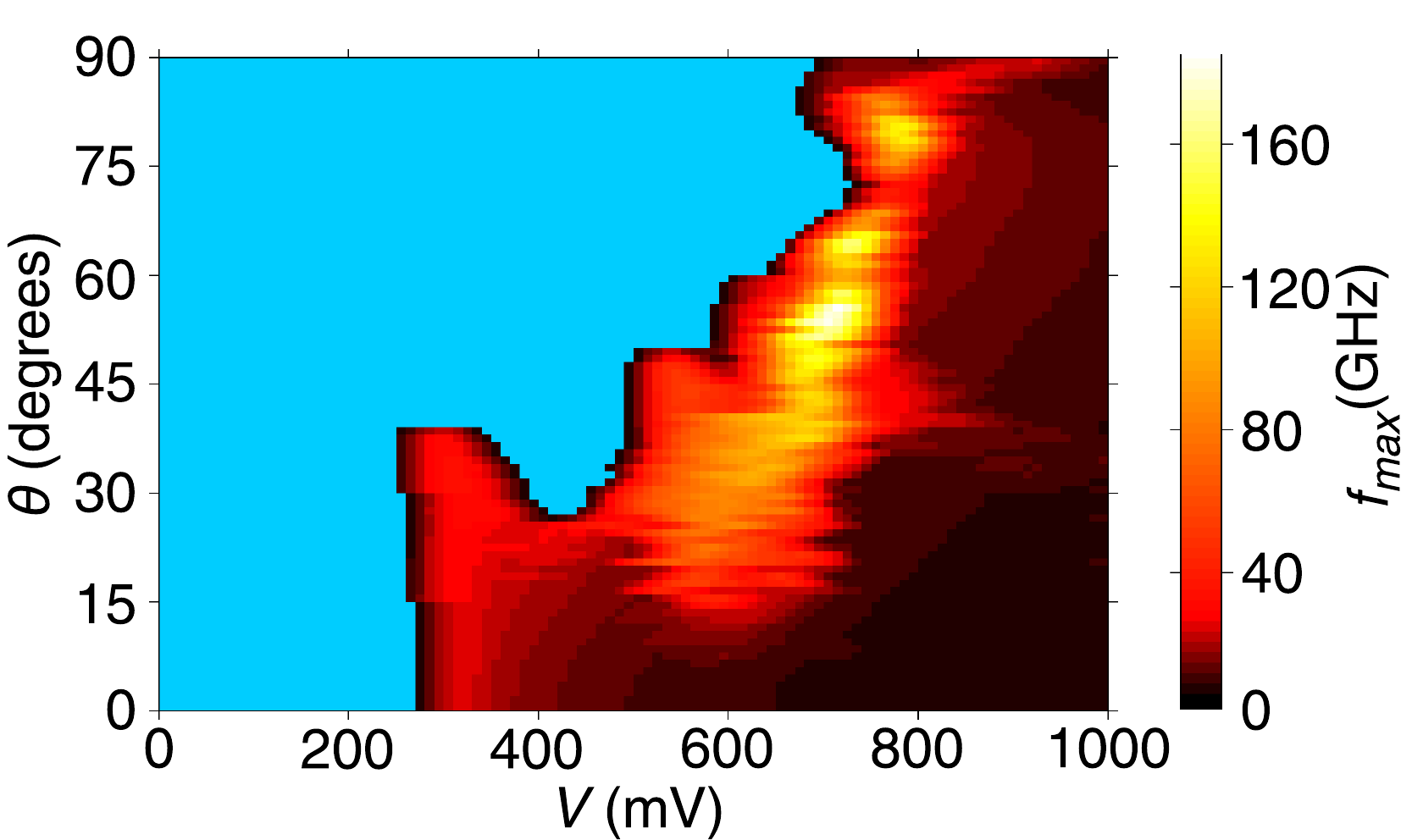}
  \caption{(Color) Color map of $f_{max}(V, \theta)$ for $B$ = 15 T. \label{fig:FP}}
\end{figure}

%To quantify the total power of the high-frequency $I(t)$ oscillations, we calculate $P_{int}=\int P(f) df$, where the integral is over $f>$ 0.2 THz. Figure \ref{fig:FP} shows a color map of $P_{int}$ in the $V- \theta$ plane. For $V<V_c$ (left of dashed curve in Fig. 5),  $P_{int}=0$ because there are no charge domain oscillations (see also Fig. 2). When $V>V_c$, where $I(t)$ oscillations do occur, $P_{int}$ generally increases with increasing $\theta$, attaining a maximum (dark red area) when $V \approx 800$ mV and $\theta \approx 70^{\circ}$. In this regime, $P_{int}$ is an order of magnitude higher than for $\theta=0^{\circ}$ due to the formation of multiple propagating charge domains.

To quantify the effect of chaos-assisted transport on the $I(t)$ oscillations, the color map in Fig. \ref{fig:FP} shows the frequency, $f_{max}$, corresponding to the largest peak in $P(f)$, in the $V- \theta$ plane. For $V<V_c$ (blue area in Fig. \ref{fig:FP}), there are no charge domain oscillations. When $V>V_c$, where $I(t)$ oscillations do occur, $f_{max}$ generally increases with increasing $\theta$, attaining a maximum (light yellow area) when $V \approx 800$ mV and $\theta \approx 70^{\circ}$. In this regime, $f_{max}$ is $\sim10$ times higher than for $\theta=0^{\circ}$ due to the formation of multiple propagating charge domains.

In conclusion, both the form and dynamics of traveling charge domains in a biased SL can be controlled and strongly enhanced by applying a tilted \textbf{B}-field. Additional NDV regions created by \textbf{B} induce multiple charge domains, which increase both the amplitude and frequency of the oscillations in $I(t)$ -- giving an order of magnitude increase in the frequency and power of the dominant Fourier peak at large $\theta$. Multiple $v_d$ maxima can be created in other ways, for example by an AC electric field \cite{HYA2009}. Our results thus open routes to controlling the \emph{collective} dynamics of charge domains in SLs by using \emph{single-electron} miniband transport to tailor $v_d(F)$. It may also be possible to realize and exploit related dynamics in nonlinear atomic and optical systems \cite{SCO02,WIL03,WIL01}.

We thank A. Patan\`e and L. Eaves for helpful discussions. This work is supported by EPSRC.


\begin{thebibliography}{}
\bibitem{footA}{We only consider the $x$-component of the electron speed because current is measured along the $x-$direction in experiment \cite{FRO2004,FOW2007}.}
\end{thebibliography}
\end{document}